\documentclass[%
 reprint,
 superscriptaddress,
 amsmath,amssymb,
 aps,
 pre,
 onecolumn,
longbibliography
]{revtex4-2}

\usepackage{graphicx}
\usepackage{dcolumn}
\usepackage{bm}
\usepackage{upgreek}
\usepackage{bm}
\usepackage[dvipsnames]{xcolor}
\usepackage{todonotes}

\usepackage{hyperref}
\usepackage[capitalise]{cleveref}

\begin{document}

\preprint{APS/123-QED}


\title{How elasticity affects bubble pinch-off}

\author{Coen I. Verschuur}
\affiliation{Physics of fluids department, University of Twente, 7500 AE Enschede, The Netherlands}

\author{Alexandros T. Oratis}
\affiliation{Physics of fluids department, University of Twente, 7500 AE Enschede, The Netherlands}

\author{Vatsal Sanjay}
\affiliation{Physics of fluids department, University of Twente, 7500 AE Enschede, The Netherlands}
\affiliation{%
CoMPhy Lab, Department of Physics, Durham University, Science Laboratories, South Road, Durham DH1 3LE, United Kingdom.
}%

\author{Jacco H. Snoeijer}
\email[]{j.h.snoeijer@utwente.nl}
\affiliation{Physics of fluids department, University of Twente, 7500 AE Enschede, The Netherlands}

\date{\today}

\begin{abstract}
The pinch-off of bubbles in viscoelastic liquids is a fundamental process that has received little attention compared to viscoelastic drop pinch-off.
While these processes exhibit qualitative similarities, the dynamics of the pinch-off process are significantly different.
When a drop of a dilute polymer solution pinches off, a thread is known to develop that prevents breakup due to the diverging polymer stresses.
Conversely, our experiments reveal that this thread is absent for bubble pinch-off in dilute polymer solutions.
We show that a thread becomes apparent only for high polymer concentrations, where the pinch-off dynamics become very sensitive to the size of the needle from which the bubble detaches.
The experiments are complemented by numerical simulations and analytical modeling using the Oldroyd-B model, which capture the dilute regime.
The model shows that polymer stresses are still singular during bubble pinch-off, but the divergence is much weaker as compared to drop pinch-off.
This explains why, in contrast to droplets, viscoelastic bubble-threads do not appear for dilute suspensions but require large polymer concentrations.
\end{abstract}

\maketitle


\section{Introduction}
\label{sec:introduction}
The formation of drops and bubbles involves the phenomenon of pinch-off, during which a body of fluid separates into two bodies \cite{Eggers.Fontelos2015}.
This transition constitutes a topological change, and causes a singularity characterized by vanishing length scales and diverging velocity and stress.
This phenomenon is observed for both liquids and gases, which is referred to as drop and bubble pinch-off respectively \cite{Papageorgiou1995, Day.etal1998, Eggers1997, Burton.etal2005, Gordillo.etal2005, Keim.etal2006}.
In both cases, the narrowing region of the fluid bridge is called the neck, whose minimum width \(h\) decreases with time \(t\).
Although one might expect only minor differences between drops and bubbles, the physical forces driving the pinch-off dynamics differ significantly.
In the inviscid limit, the minimal neck width for drop pinch-off scales as \(h \propto (t_0 - t)^{2/3}\), emerging from the balance of capillarity and inertia \cite{Eggers.Fontelos2015, Day.etal1998}.
For inviscid bubble pinch-off, by contrast, surface tension is sub-dominant during the final stages prior to breakup; a purely inertial balance emerges, characterized by an effective dynamics \( h \propto (t_0 - t)^{\alpha} \) where the effective exponent is 1/2 up to logarithmic corrections \cite{Eggers.Fontelos2015} (typical reported values \(\alpha \approx 0.56\) \cite{Burton.etal2005, Gordillo.etal2005, Keim.etal2006, Bergmann.etal2006, Eggers.etal2007, Gekle.etal2009, Thoroddsen.etal2007}).
A consequence is that bubble pinch-off is a bit faster, and thereby more difficult to measure experimentally.

Many applications make use of viscoelastic liquids rather than Newtonian liquids, examples of which can be found in inkjet printing \cite{Morrison.Harlen2010, Sen.etal2021} and spraying \cite{Makhnenko.etal2021,Gaillard.etal2022a} among others.
Viscoelastic fluids typically contain some long polymers that give the fluid elastic properties.
Indeed, when the fluid is deformed, the polymers are stretched, resulting in an elastic restoring force \cite{Snoeijer.etal2020, Morozov.Spagnolie2015, Larson1999, Tanner2000}.
These viscoelastic properties of the fluid have a significant influence on the pinch-off process, as is well-known for drops \cite{Clasen.etal2006, Bazilevsky.etal1990, Entov.Hinch1997, Anna.McKinley2001, Amarouchene.etal2001, Eggers.etal2020, Deblais.etal2020}.
Even for very low concentrations of polymer, as small as \(0.01\) wt\%, the drop pinch-off is dramatically delayed \cite{Clasen.etal2006}.
Typically, the dynamics can be split into an initial Newtonian regime that is followed by a late-time viscoelastic regime.
In the viscoelastic regime, the polymeric stress prevents the neck from pinching off immediately, and so the relaxation of the polymers limits the thinning of the neck.
For dilute polymer solutions, the Oldroyd-B model predicts the formation of a long thread that is thinning exponentially with a decay rate of \(3 \lambda\), where \(\lambda\) is the relaxation time of the polymer solution \cite{Bazilevsky.etal1990, Bazilevskii.etal1997, Entov.Hinch1997}.
As such, the pinch-off of drops has often been used as a rheological tool, to characterize the relaxation time of polymer solutions \cite{Anna.McKinley2001}, though the experimentally-measured relaxation time can be susceptible to the pinch-off geometry \cite{Gaillard.etal2023}.

In spite of its potential importance, and in contrast to droplet breakup, only a few studies are dedicated to the pinch-off of bubbles in viscoelastic fluids \cite{Jiang.etal2017,Rajesh.etal2022}.
For low-viscosity polymer solutions, the phenomenology at first sight bears a strong resemblance to the case of drop breakup.
The initial dynamics resembles that of bubbles in Newtonian liquids.
Subsequently, viscoelasticity slows the dynamics without altering the overall neck shape, resulting once again in an elongated air thread \cite{Jiang.etal2017, Rajesh.etal2022}.
Rajesh et al. \cite{Rajesh.etal2022} even found a hint of an exponential thinning.
However, the tenuous air-threads were much thinner than those observed during drop breakup.
In addition, the characteristic timescale for the thinning of the air-thread turned out to be much smaller than the relaxation time, so that the accurate quantification remains a challenge.

\begin{figure}[tpb]
	\centering
	\includegraphics[width=0.75\linewidth]{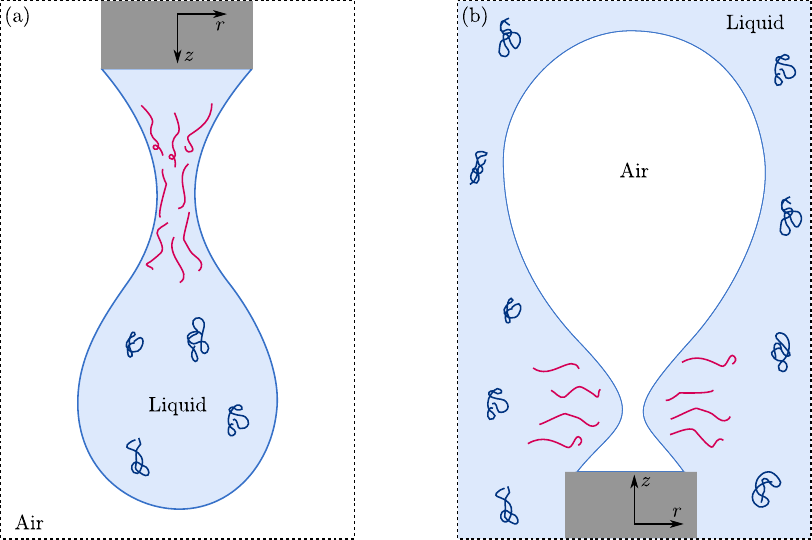}
  \caption{Cartoon of viscoelastic drop and bubble pinch-off. (a) When a drop pinches off, the polymers become stretched in the longitudinal direction \(z\), following the stretching of the neck \cite{ingremeau2013stretching}. (b) When a bubble pinches off, the neck is still stretched in the longitudinal direction, but the polymers are stretched in the radial direction \(r\). Note that the polymers are not drawn to scale.}
	\label{fig:schematic_pinch-off}
\end{figure}

From a theoretical point of view, it is not clear whether and how viscoelastic bubble pinch-off is related to that of drops.
Indeed, major differences in polymer stress were found to emerge in \emph{coalescence} of drops and bubbles \cite{Oratis.etal2023}, owing to different flow fields \cite{Oratis.etal2023}.
To illustrate this, we sketch the expected configuration of polymers during breakup in \cref{fig:schematic_pinch-off}.
The predominant stretching of polymers, as expected from the kinematics of the flow and confirmed by the numerical simulations presented later in the manuscript (\cref{sec:Model}), occurs along different directions: along the neck for drops (a) and towards the neck for bubbles (b) (see \cref{app:flow}).
We thus anticipate the dominant stress along the \(z\) direction for drops, and along the $r$ direction for bubbles.
While it is clear that the viscoelastic stress is expected to slow down bubble pinch-off, a quantitative analysis is currently lacking.

In this paper we aim to quantify the pinch-off dynamics of bubbles in viscoelastic liquid, using experiments dedicated to measuring the tiny neck of air.
Specifically, our goal is to characterize the pinching of the air-neck by capturing the onset of its formation, its thinning dynamics, and its eventual break-up.
Besides varying different polymer lengths and concentration, we also consider the influence of the needle size (motivated by recent work on viscoelastic drops \cite{Gaillard.etal2023}).
In addition to the experiments, we will also perform numerical simulations using the Oldroyd-B fluid, which should capture the dilute regime, and develop an analytical model that faithfully describes the elastic stress.
\cref{sec:Experiments} presents the experimental and numerical methods.
The experimental results are found in \cref{sec:results}, followed by the modeling in \cref{sec:Model}.
The paper closes with a conclusion in \cref{sec:Conclusion}.

\section{Methods}
\label{sec:Experiments}
\subsection{Experimental}
The bubble pinch-off experiments are performed in a custom-made acrylic container (\(3\times 3 \times 6\,\)cm) with a needle (Nordson, general purpose dispense tip) at the bottom.
The needle is connected to a syringe pump (Harvard Apparatus, PHD 2000), which injects air at a quasistatic flow rate.
The bubble pinch-off process is recorded using back light illumination (Schott KL2500) and a high-speed camera (Photron SA-Z type 2100k) at a frame rate of \(400\,000\) fps.
A microscope with a long working distance \(20 \times\) objective (Olympus SLMPlan N 20×/0.25) records the bubble pinch-off process, resulting in a resolution of 1.0 \textmu m/pixel.
The contour of the bubble neck is extracted from the images using custom image processing techniques based on maximum intensity gradient detection.
Using those contours, we can determine the minimal width of the neck as a function of time.

To study the effects of viscoelasticity on the pinch-off dynamics we use viscoelastic polymer solutions of polyethylene oxide (PEO, \(M_W = 2.0 \times 10^6\) and \(4.0 \times 10^6\, \textrm{g/mol}\), Sigma-Aldrich).
The polymer solutions were prepared by making a stock solution of \(1\,wt.\,\%\) of polymer dissolved in ultra-pure water (MilliQ, Millipore Corporation) at room temperature and stirring the solution on a roller mixer (Stuart, SRT6D) until the polymer was fully dissolved after 5 days.
The lower concentrations were made by diluting the stock solution and stirring the solution again on the roller mixer.
For the experiments we changed the polymer length (\(M_W = 2.0 \times 10^6\) and \(4.0 \times 10^6 \textrm{g/mol}\)) the polymer concentration (\(0 - 1\,wt.\,\%\)) and the needle diameter (\(0.41 - 1.54\,\textrm{mm}\)).
We conducted all the experiments from low to high concentrations to minimize the effect of improper cleaning of the container.
To avoid degradation of polymer solutions, all measurements were done within one week of preparation.
Furthermore, for every set of measurements new fluid of the same batch was taken for every 5 measurements, since the violent collapse of the neck could break the polymer chains in the solution.

The viscosity of the polymer solutions is measured using a rheometer (Anton Paar, MCR 502) with a cone-plate geometry (CP50-1).
All the measurements are performed at room temperature, and the shear rate is increased from \(0.001 - 1000\, \textrm{s}^{-1}\) (see \cref{app:rheology}).
Furthermore, an overlap concentration (\(c^*\)) is defined, such that the bulk concentration is equal to the concentration of one single polymer coil \cite{Clasen.etal2006, deGennes1979}.
While the overlap concentration isn't strictly defined, we use an overlap concentration of \(c^* \approx 0.7\,wt.\,\%\) and \(c^* \approx 0.99\,wt.\,\%\), for PEO, \(M_W = 2.0 \times 10^6\) and \(4.0 \times 10^6\, \textrm{g/mol}\) respectively \cite{Dekker.etal2022, Kawaguchi.etal1997}.
Finally, the effective relaxation time (\(\lambda_d\)) of the polymer solution is determined with the drop pinch-off experiment by measuring the width of the neck over time and fitting the data with the Oldroyd-B model (see \cref{app:relaxation_time}).
This is in line with \cite{Bazilevsky.etal1990, Entov.Hinch1997, Anna.McKinley2001, Bazilevskii.etal1997, Gaillard.etal2023} that use thinning of jets to characterize the effective relaxation time of the solution.
Owing to the polydispersity (polydispersity index of \(\sim2\)) of high molecular weight PEO \cite{Stokes1998}, this measured relaxation time should be interpreted as an effective timescale that represents a spectrum of relaxation times rather than a single molecular timescale.

\subsection{Numerical}
\label{subsec:numerical}

\subsubsection{Model configuration}

We study the breakup dynamics of both air and liquid necks in viscoelastic media using idealized model configurations. Rather than simulating complete bubble or drop pinch-off processes, we consider initially cylindrical necks with imposed sinusoidal perturbations and investigate their capillary-driven breakup. This approach isolates the fundamental physics of neck thinning in viscoelastic media while avoiding the geometric complexities of the pinch-off near needles or nozzles.

We examine two complementary configurations:
\begin{itemize}
\item \textbf{Air neck breakup}: A cylindrical air filament of radius \(h_0\) embedded in a viscoelastic liquid, modeling the final stages of bubble pinch-off where the surrounding medium exhibits viscoelastic properties.
\item \textbf{Liquid neck breakup}: A cylindrical viscoelastic liquid filament of radius \(h_0\) surrounded by Newtonian gas, representing the drop pinch-off counterpart.
\end{itemize}
Both configurations employ axisymmetric domains with periodic boundary conditions, allowing us to track a single perturbation wavelength. We emphasize that while these model systems capture essential breakup physics, they represent idealizations that differ from complete bubble/drop pinch-off geometries.

\subsubsection{Governing equations}
The mass and momentum conservation equations for both phases read:

\begin{align}
\nabla \cdot \mathbf{u} &= 0, \label{eq:continuity_num}\\
\rho \left(\frac{\partial \mathbf{u}}{\partial t} + \nabla \cdot (\mathbf{u}\mathbf{u})\right) &= -\nabla p + \nabla \cdot \boldsymbol{\tau} + \mathbf{f}_\gamma, \label{eq:momentum_num}
\end{align}

\noindent where \(\boldsymbol{\tau}\) represents the stress tensor and \(\mathbf{f}_\gamma\) the force due to surface tension \(\gamma\).
The product \(\mathbf{u}\mathbf{u}\) denotes the dyadic product, so that \(\nabla\cdot(\mathbf{u}\mathbf{u})\) is the conservative-flux form of the inertial advection. For incompressible flow, \(\nabla\cdot\mathbf{u}=0\), this is equivalent to the more familiar \(\mathbf{u}\cdot\nabla\mathbf{u}\) \cite{Popinet2009}.
We include surface tension using the continuum-surface-force formulation. In the sharp-interface description, the surface-tension force is a singular force density
\(
\mathbf{f}_\gamma = \kappa \delta_s \mathbf{n},
\)
where \(\delta_s\) is a surface delta distribution supported on the interface and \(\mathbf{n}\) is the normal to the interface. If \(C\) is the exact phase-indicator function (i.e., a Heaviside function across the interface), then
\(
\nabla C = \delta_s \mathbf{n}
\)
in the distributional sense. In the numerical VoF method employed here, however, \(C\) is stored as a cell-averaged volume fraction and the interface is reconstructed over a finite grid cell. Consequently, the singular quantity \(\delta_s\mathbf{n}\) is represented by the regularised/discrete gradient of \(C\), giving
\(
\mathbf{f}_\gamma \approx \kappa \nabla C .
\)
Here \(\kappa\) is computed from the VoF interface using height functions \cite{Brackbill.etal1992}.

In the viscoelastic phase (liquid pool for air necks, liquid neck for drop case), we decompose the stress as:
\begin{align}
\boldsymbol{\tau} = \boldsymbol{\tau}_s + \boldsymbol{\tau}_p, \label{eq:stress_decomp}
\end{align}

\noindent with the solvent contribution:
\begin{align}
\boldsymbol{\tau}_s = 2\eta_s\mathbf{D}, \label{eq:solvent_stress_num}
\end{align}

\noindent where \(\mathbf{D} = (\nabla\mathbf{u} + \nabla\mathbf{u}^T)/2\) is the symmetric part of the velocity gradient tensor and \(\eta_s\) represents the solvent viscosity.
The polymeric stress follows the Oldroyd-B model:
\begin{align}
\boldsymbol{\tau}_p = G(\mathbf{A} - \mathbf{I}), \label{eq:polymer_stress_num}
\end{align}
where \(G\) is the elastic modulus.
The Oldroyd-B model is used here as it is the simplest model that captures the essential viscoelastic behavior for drop pinch-off.
The conformation tensor evolves according to:
\begin{align}
\frac{\partial \mathbf{A}}{\partial t} + \mathbf{u} \cdot \nabla \mathbf{A} - \mathbf{A} \cdot \nabla \mathbf{u} - (\nabla \mathbf{u})^T \cdot \mathbf{A} = -\frac{1}{\lambda}(\mathbf{A} - \mathbf{I}), \label{eq:conformation_num}
\end{align}

\noindent where \(\lambda\) represents the relaxation time.

While we have the governing equations in dimensional form here, we use the dimensionless forms for the numerical implementation by using the initial neck radius \(h_0\) as the characteristic length scale, the inertio-capillary timescale \(\tau_{ic} = \sqrt{\rho_lh_0^3/\gamma}\), and \(\gamma/h_0\) as the pressure scale.
This results in the following dimensionless numbers: the Ohnesorge number \(Oh = \eta_s/\sqrt{\rho_l \gamma h_0}\) (viscosity), the Deborah number \(De = \lambda/\tau_{ic}\) (relaxation time), and the elastocapillary number \(Ec = G h_0/\gamma\) (elasticity).

Throughout this work, we take the limiting case of \(De \to \infty\) where the polymeric medium has a perfect memory of the flow \cite{dixit2025viscoelastic, Oratis.etal2024}.
In this limit there is no relaxation of stress, which thus generates the maximum elastic stress possible for the Oldroyd-B fluid \cite{Snoeijer.etal2020}.

\subsubsection{Numerical implementation}

We implement these equations in Basilisk C \citep{basilliskpopinet} using the volume-of-fluid method for interface capturing \cite{sanjayComphylabMultiRheoFlowV012025,sanjayComphylabElasticPinchOffInitial2025}.
The viscoelastic constitutive equations employ the log-conformation method \citep{fattal2004constitutive,lopez2019adaptive} to ensure numerical stability at high Deborah numbers.
For the elastic limit (\(De \to \infty\)), we utilize a modified formulation that enforces the affine motion constraint.

Surface tension acts through the continuum surface force formulation:
\begin{align}
\mathbf{f}_\gamma \approx \kappa \nabla C, \label{eq:csf_num}
\end{align}

\noindent where \(\kappa\) denotes the interface curvature computed via height functions, and \(C\) the VoF color function.
We initialize both neck configurations with a sinusoidal perturbation:
\begin{align}
h(z,t=0) = h_0[1 + \epsilon \sin(kz)], \label{eq:perturbation_num}
\end{align}

\noindent where \(\epsilon = 0.05\) represents the perturbation amplitude and \(k = 1\) the wavenumber, yielding a wavelength \(\lambda_p = 2\pi h_0\). The computational domain spans one full wavelength (\(L_0 = 2\pi\) in dimensionless units) with periodic boundary conditions in the axial direction.
We apply axisymmetric conditions along \(r = 0\) and extend the domain radially to minimize boundary effects.

We employ adaptive mesh refinement based on the VoF function, velocity components, conformation tensor components, and interface curvature, with error tolerances of \(10^{-3}\), \(10^{-2}\), \(10^{-2}\), and \(10^{-6}\), respectively.
The maximum refinement level yields minimum grid sizes from \(\Delta = h_0/512\) to \(\Delta = h_0/2048\) as needed for thin necks.

For both configurations, we maintain density and viscosity ratios of \(\rho_g/\rho_l = 10^{-3}\) and \(\eta_g/\eta_l = 10^{-2}\) between gas and liquid phases. We explore the parameter space spanning \(Oh = 10^{-2}\) unless otherwise stated, \(Ec = 10^{-2}\) to \(4 \times 10^{-1}\) in the elastic limit (\(De \to \infty\)).

This model approach, while simplified compared to actual bubble and drop pinch-off, enables systematic investigation of how viscoelasticity fundamentally alters neck breakup across the full range of relevant parameters.

\begin{figure}[tpb]
	\centering
	\includegraphics[width=17.2cm]{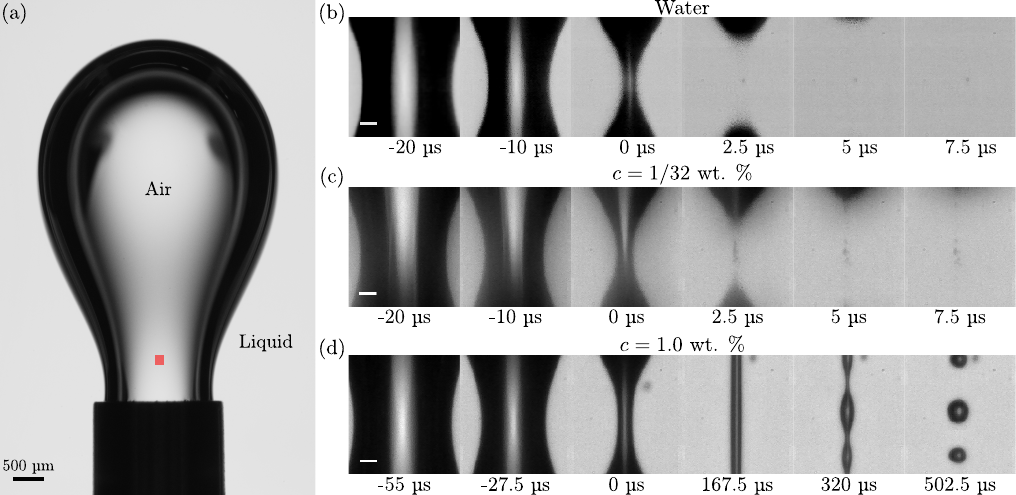}
  \caption{Snapshots of the bubble pinch-off process for different PEO concentrations (\(M_W = 4.0 \times 10^6 \, \textrm{g/mol}\)). (a) Overview image of the bubble pinch-off process, where the red square roughly indicates the region in which pinch-off occurs. (b), (c) and (d) show the bubble pinch-off process for a polymer concentration of \(0\)\, \textrm{wt.}\, \%, \(1/32\, \textrm{wt.}\, \%\) and \(1\, \textrm{wt.}\, \%\), respectively. The scale bar in (b), (c) and (d) is \(20\) \textmu m. (supplemental Movie S1-3 respectively \cite{supplemental_material}).}
	\label{fig:bubble_seq}
\end{figure}

\section{Results}
\label{sec:results}
\subsection{Phenomenology}
The bubble pinch-off process for different concentrations of PEO, \(M_W = 4.0 \times 10^6 \, \textrm{g/mol}\) is illustrated in \cref{fig:bubble_seq}.
When the bubble grows, the buoyancy force pushes it up, causing a neck to form.
This neck violently collapses, resulting in the bubble pinching off.
The shape of the initial collapse is similar for water and the two polymer concentrations (compare the first panels of b, c and d).
The difference between the solutions becomes apparent only after the transition from Newtonian to viscoelastic, at \(t-t_0=0\), where \(t_0\) marks the moment pinch-off would occur for a Newtonian fluid. Specifically, for the high polymer concentration in \cref{fig:bubble_seq}d we clearly observe the formation of an air-thread. The following section will give a more precise definition of \(t_0\) for polymer solutions; we will refer to ``thread" as the thin structure that forms after $t_0$.

\begin{figure}[tpb]
  \centering
  \includegraphics[width=\linewidth]{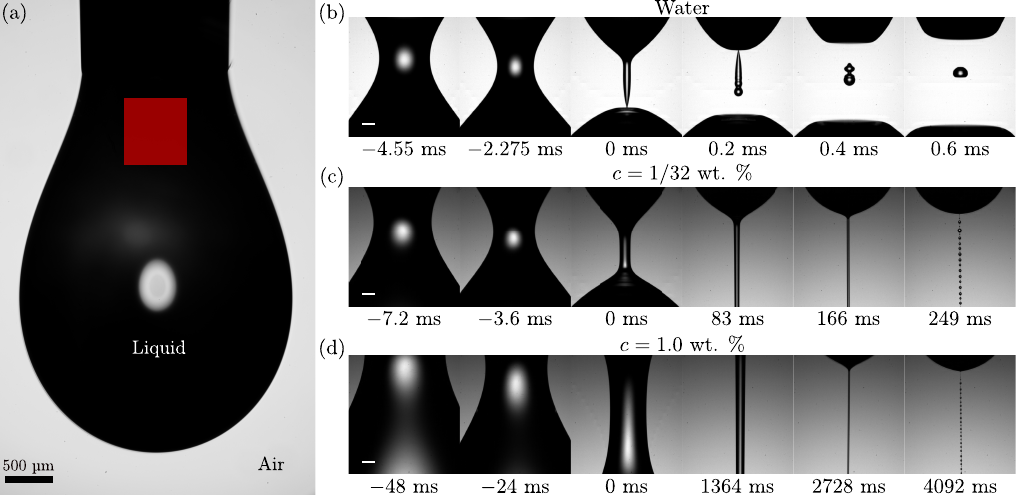}
  \caption{Snapshots of the drop pinch-off process for different PEO concentrations (\(M_W = 4.0 \times 10^6 \, \textrm{g/mol}\)). (a) Overview image of the drop pinch-off process, where the red square indicates roughly the region where the images in (b), (c) and (d) are taken. (b), (c) and (d) show the drop pinch-off process for a polymer concentration of \(0\, \textrm{wt.}\%\), \(1/32\, \textrm{wt.}\%\) and \(1\, \textrm{wt.}\%\), respectively. The scale bar in (b), (c) and (d) is \(250\) \textmu m. (supplemental Movie S4-6 respectively \cite{supplemental_material}).}
  \label{fig:drop_seq}
\end{figure}

The natural counterpart of the bubble pinch-off is the drop pinch-off, which is shown in \cref{fig:drop_seq}.
The drop pinch-off process is initiated by gravity that pulls the droplet downwards, causing a neck to form.
The initial dynamics of the neck can be described by balancing the capillary pressure (\(\gamma/h\)) by the inertia (\(\rho(h/t)^2\)), as discussed before.
For the polymer solutions (panels c and d), we see a transition from the Newtonian regime to the viscoelastic regime, where a thread is formed and, in the end, the well-known beads-on-a-string phenomena.

When comparing the two pinch-off processes, we see some similarities and important differences, including the thread's time scale and overall width, in line with \cite{Rajesh.etal2022}.
In drop pinch-off, the initial thread width is approximately \(1/10\textsuperscript{th}\) of the needle width, while for bubble pinch-off, we see that the initial thread width is closer to \(1/100\textsuperscript{th}\) of the needle width.
Furthermore, the thread in drop pinch-off is visible for a much longer time than the thread in bubble pinch-off.
For bubble pinch-off, a concentration of \(1/32\, \textrm{wt.} \, \%\) is already too low for a visible thread to emerge, while for drop pinch-off a concentration as small as \(1/100\, \textrm{wt.} \, \%\) would be sufficient to see the effect of the polymers.
It could be that the bubble thread does not form, or that the spatio-temporal resolution (which is improved with respect to \cite{Rajesh.etal2022}) is still not sufficient to reveal the thread for dilute suspensions.
Either way, it is clear that drop and bubble pinch-off are significantly different, even though they look similar at first sight.
So, in our measurements, we focus on the collapse of the neck and the effect of the polymer concentration and the needle size, which is further explored in the following sections.

\subsection{Effect of concentration and needle size}
\label{subsec:concentration_and_needle}

\begin{figure}[tpb]
	\centering
	\includegraphics[width=17.2cm]{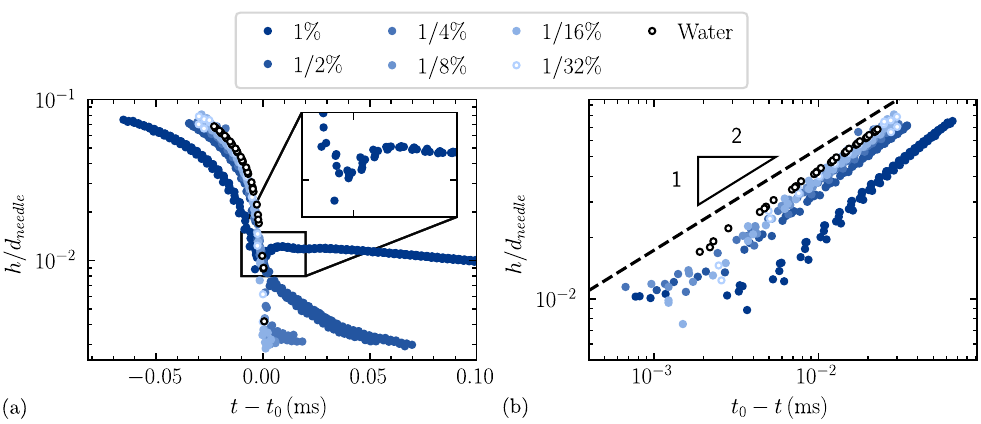}
  \caption{The width of the neck at the center is normalized by the diameter of the needle (\(d_{needle} = 1.54\) mm) over time (\(t-t_0\)) for different concentrations of PEO (\(M_W = 4.0 \times 10^6 \, \textrm{g/mol}\)).}
	\label{fig:concentration}
\end{figure}

We now turn to a quantitative analysis of the pinching dynamics.
The minimal width of the neck \(h\) normalized by the diameter of the needle \(d\) against time \(t-t_0\) is plotted in \cref{fig:concentration} for different concentrations of PEO, for the case of \(M_W = 4.0 \times 10^6 \, \textrm{g/mol}\).
We define \(t_0\) as the moment in which the neck would pinch-off for a Newtonian fluid \cite{Dekker.etal2022}.
To be precise, this time is determined by fitting the data close before the pinch-off with the scaling law \(A\left(t_0 - t\right)^\alpha\).
Indeed, such a power-law dynamics is observed prior to thread formation, as is shown in \cref{fig:concentration}b, with \(\alpha \approx 0.56-0.6\) \cite{Thoroddsen.etal2007, Eggers.etal2007}.
In \cref{fig:concentration}a, we can distinguish two different regimes, the Newtonian regime (\(t-t_0 < 0\)) and the viscoelastic regime (\(t-t_0>0)\).
In the Newtonian regime, we see a slight difference between the concentrations, which is mainly in the prefactor of the scaling law.
This difference in prefactor is similar to what is seen by bubble pinch-off in Newtonian fluids \cite{Burton.etal2005, Thoroddsen.etal2007}, where it was found that a higher viscosity results in a lower prefactor in the scaling law.

Based on the shape and the transition from the Newtonian to the viscoelastic regime, one could be tempted to fit an exponential decay in the viscoelastic regime, such that the resulting decay rate $\lambda_b$ can be directly compared  with the drop relaxation time $\lambda_d$.
Because the slope is significantly steeper in the viscoelastic regime of bubbles as compared to the equivalent one of drops, such a fit would result in \(\lambda_b \ll \lambda_d\).
However, the few cases that exhibit a thread have a limited range of data points available in the viscoelastic regime.
Therefore, it becomes questionable to assume that the thread thins exponentially.

\begin{figure}[tpb]
	\centering
	\includegraphics[width=17.2cm]{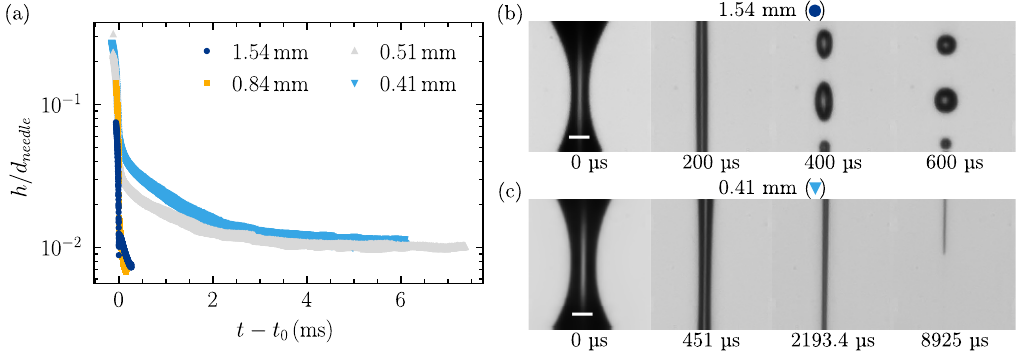}
  \caption{The effect of needle diameter \(d_{needle}\). (a) The normalized width of the neck \(h/d_{needle}\) plotted against time (\(t-t_0\)) at a concentration of \(1\, \textrm{wt.} \, \%\) PEO (\(M_W = 4.0 \times 10^6 \, \textrm{g/mol}\)). (b) For the largest needle size used \(d_{needle} = 1.54\) mm, the thread disintegrates into smaller bubbles over a period of approximately 400 \textmu s.
  (c) For the smallest needle size used \(d_{needle} = 0.41\) mm, the thread ruptures after milliseconds and retracts towards the bubble. The scale bar in (b) and (c) is \(20\) \textmu m. (supplemental Movie S7 and S8 respectively \cite{supplemental_material}).}
	\label{fig:needle_size}
\end{figure}

\begin{figure}[tpb]
	\centering
	\includegraphics[width=\linewidth]{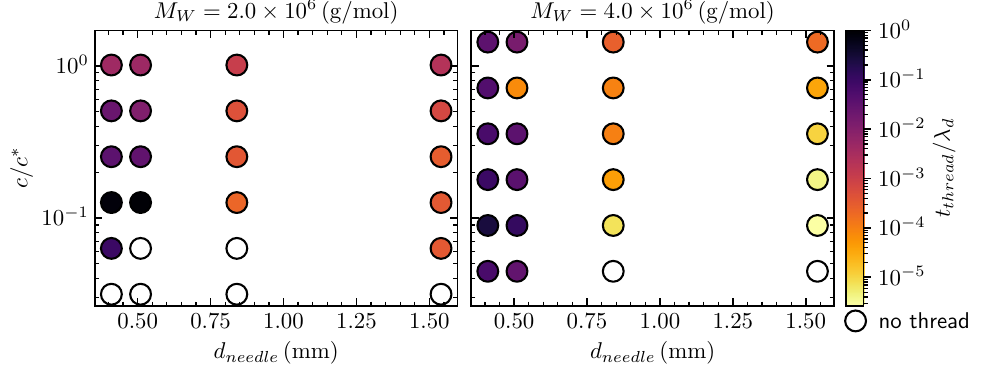}
	\caption{The thread duration $t_{thread}$ normalized by the relaxation time as function of the normalized polymer concentration $c/c^*$ and the needle diameter $d_{needle}$. The left shows the results of PEO, \(M_W = 2.0 \times 10^6 \, \textrm{g/mol}\), and right shows the results of PEO, \(M_W = 4.0 \times 10^6 \, \textrm{g/mol}\).}
	\label{fig:thread_duration}
\end{figure}

The ambiguity of an exponential thinning becomes even more evident when we extend our results to smaller needle sizes, as shown in \cref{fig:needle_size}a.
This plot also shows the minimal width of the neck \(h\) normalized by the diameter of the needle \(d_{needle}\) against time \(t-t_0\), but now for different needle sizes at the same polymer concentration of \(1\, \textrm{wt.} \, \%\) PEO, \(M_W = 4.0 \times 10^6 \, \textrm{g/mol}\).
We note that the dark blue data points in \cref{fig:concentration}a and \cref{fig:needle_size}a are the same dataset, and these could be consistent with an exponential decay of the neck width. However, the duration of the thread is too short to make a firm quantitative conclusion. In contrast, for smaller needle sizes, the neck thinning lasts much longer and is not compatible with an exponential decay. Instead, the curve flattens, approaching a nearly constant width. As a result, the thickness decreases much more slowly than would be expected for an exponential decay.

In addition to the difference in thread duration, we also notice a difference in the spatial structure during the thread breakup, as can be seen in \cref{fig:needle_size}b and c.
For large needle sizes (\cref{fig:needle_size}b), the thread breaks up into multiple satellite bubbles, reminiscent of the Rayleigh-Plateau instability of a liquid jet.
This unstable breakup likely occurs due to an instability induced during the transition from the Newtonian to the viscoelastic regime.
During this transition towards the viscoelastic regime, the thread width suddenly increases in size during its collapse.
This rapid rebound suggests that the thread gets locally compressed, resulting in an instability that disintegrates the thread into multiple bubbles, leading to a short thread duration.
For these larger needle sizes we also see that the thread can lose its axisymmetry at the final stages before breakup. An example of this is provided in supplemental Movie S9 \cite{supplemental_material}.
For small needle sizes, in contrast, the thread ruptures at the bottom near the needle and retracts upwards towards the bubble (\cref{fig:needle_size}c), reminiscent of Taylor-Culick flow.
The compression that we observe for the large needle sizes is absent for smaller needle sizes, where the thread continues to thin, leading to prolonged durations.
Most previous studies on bubble pinch-off typically involved larger needle sizes, resulting only in the unstable breakup of the thread.

To further investigate the effects of the polymer length, concentration, and needle size, we measured the thread duration.
We define the thread duration by the time we see a thread forming (\(t-t_0=0\)) until the thread is not visible any more, with a minimum of one frame (\(2.5\) \textmu s) where we see a connected thread.
This duration is normalized by the effective relaxation time of the polymer solution \(\lambda_d\), determined by the drop pinch-off experiment.
The results are plotted in \cref{fig:thread_duration}, as a function of the normalized concentration and needle diameter for two polymer lengths \(M_W = 2.0 \times 10^6 \, \textrm{g/mol}\) and \(M_W = 4.0 \times 10^6 \, \textrm{g/mol}\).
The position of the data points in the graph gives the concentration and needle size, while the color gives the thread duration normalized by the relaxation time.
In the case where the thread was not visible, the color is set to white.
From \cref{fig:thread_duration} we notice that there is no thread for almost all experiments at the lowest concentration.
This result marks a significant difference from drop pinch-off, for which a very dilute suspension exhibits a clear thread.
Furthermore, we see two regimes in the thread duration as a function of the needle size.
First, we see that the needle sizes of \(0.41\) and \(0.51\, \textrm{mm}\) have a higher thread duration than the needle sizes of \(0.85\) and \(1.54\, \textrm{mm}\).
This corresponds to the difference in the breakup of the thread as is shown in \cref{fig:needle_size}b and c.
Additionally, we also see a different trend.
For the needle size of \(0.41\) and \(0.51\, \textrm{mm}\) we see that a higher concentration results in a longer thread duration, while for the needle size of \(0.85\) and \(1.54\, \textrm{mm}\) the thread duration is decreasing for a higher concentration.

\section{Bubble pinch-off in Oldroyd-B fluid}
\label{sec:Model}
\subsection{Basilisk simulations}
\begin{figure}[tpb]
	\centering
	\includegraphics[width=17.2cm]{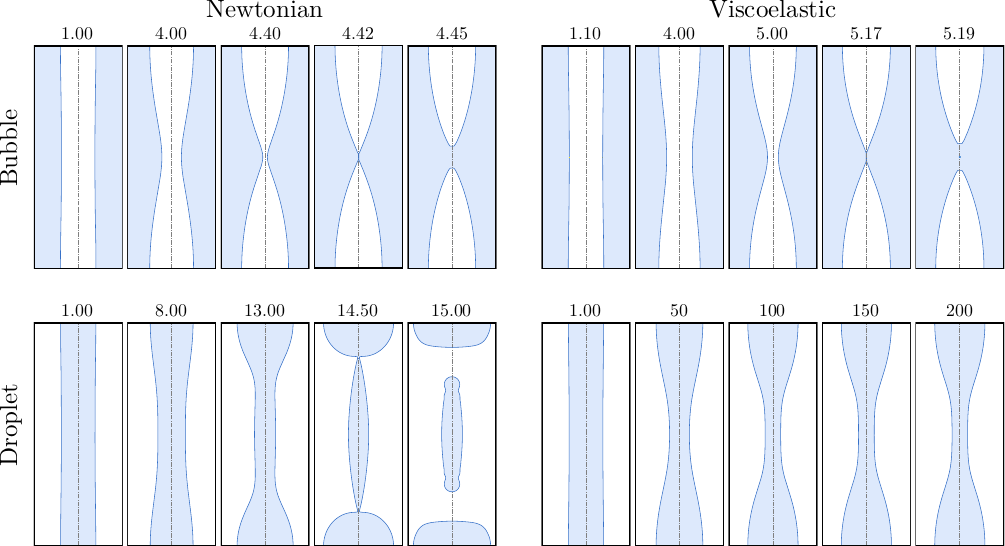}
  \caption{Snapshots of the numerical bubble and droplet pinch-off simulations in the limit of a Newtonian fluid ($Ec = 0$), and viscoelastic Oldroyd-B ($Ec = 0.1$) fluid with \(\lambda \to \infty\). The liquid is colored blue and the air phase is colored white. For all cases, $Oh = 10^{-2}$ except for the viscoelastic drop where $Oh = 1$ to eliminate inertio-capillary oscillations at low $Oh$. The timestamps are in dimensionless time given by \(\tau_{ic} = \sqrt{\rho_lh_0^3/\gamma}\).}
  \label{fig:drop_bubble_contour_simulation}
\end{figure}

To interpret the experimental observations, we conducted a series of numerical simulations.
As explained in \cref{subsec:numerical}, we adopt the Oldroyd-B model that is known to capture thread formation in viscoelastic drops and, to maximize the effect of elasticity, we impose no stress relaxation.
\cref{fig:drop_bubble_contour_simulation} shows the numerical results, comparing viscoelastic bubble pinch-off with the Newtonian case, as well as the drop pinch-off counterparts.
Let us first consider the upper panels that compare Newtonian and viscoelastic bubble pinch-off.
The air cavity evolution is very similar in both cases, with the viscoelastic dynamics being slightly slower.
This similarity is surprising since pinch-off is expected to generate diverging elastic stresses, particularly for simulations with no stress relaxation.
Yet, the elasticity does not perturb the break-up process.
The lower panels of \cref{fig:drop_bubble_contour_simulation} show the well-known cases for Newtonian and viscoelastic drops.
For drops, polymer stress clearly affects the dynamics: the initial thinning rate slows significantly, and we observe the formation of a viscoelastic thread.
While the thread normally thins exponentially, in our simulations without stress relaxation, it attains a stationary shape.

\begin{figure}[tpb]
	\centering
	\includegraphics[width=17.2cm]{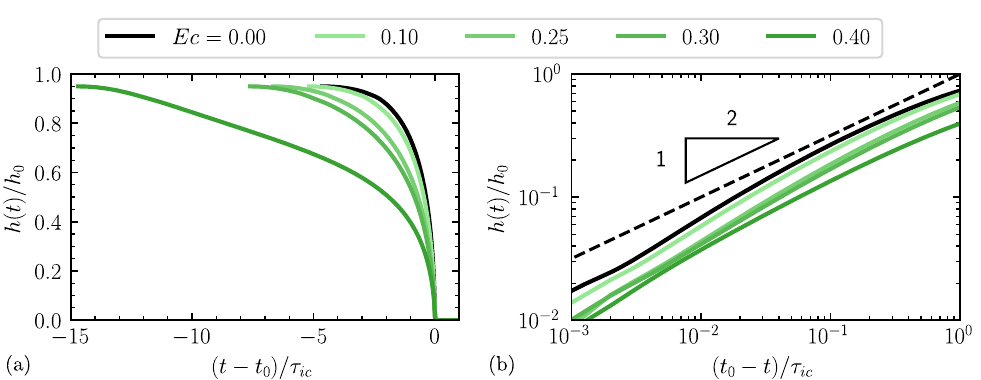}
  \caption{Comparison of the neck width for constant \(De = \infty\) and different values for the relaxation time \(Ec = 0-0.40, Oh = 10^{-2}\). a) shows the neck width as a function of \((t-t_0)/\tau_{ic}\) while b) shows the neck width as function of \((t_0-t)/\tau_{ic}\) on a log-log scale.}
  \label{fig:concentration_simulation}
\end{figure}

\begin{figure}[tpb]
  \centering
  \includegraphics[width=\textwidth]{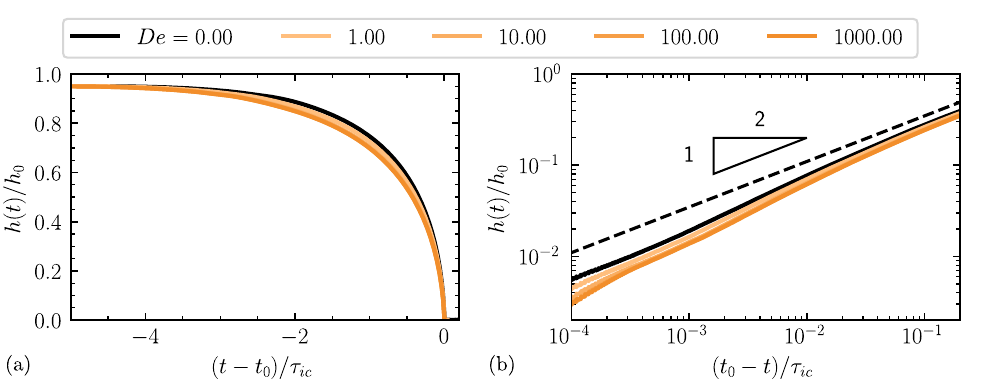}
  \caption{Comparison of the neck width for constant \(Ec = 0.1, Oh = 10^{-2}\) and different values for the relaxation time \(De = 0-1000\). a) shows the neck width as a function of \((t-t_0)/\tau_{ic}\) while b) shows the neck width as function of \((t_0-t)/\tau_{ic}\) on a log-log scale.}
  \label{fig:finite_relaxation_simulation}
\end{figure}

Qualitatively, these simulation results agree with the experimental observations of \cref{fig:bubble_seq,fig:drop_seq} for bubbles and drops, respectively.
The main difference is the formation of air threads that is observed for bubble experiments at large polymer concentrations, beyond the dilute limit.
However, the Oldroyd-B model applies only to dilute suspensions, where neither experiments nor simulations show threads.
\cref{fig:concentration_simulation}a quantifies the breakup, showing the evolution of the minimal neck width \(h(t)\) over time, as the counterpart of the experimental data in \cref{fig:concentration}.
The curves correspond to different values of \(Ec\) (the Newtonian case has \(Ec=0\)).
Once again, the simulations do not exhibit a strong sign of viscoelasticity near the moment of pinching, and \(h \to 0\) at a well-defined time \(t_0\).
The approach to the singularity is further illustrated in \cref{fig:concentration_simulation}b, showing that the thinning of the neck follows a power-law with an exponent close to \(1/2\) as we have also seen for the experimental results. For completeness, we also report simulations for finite relaxation times, characterized by different values of $De$. The results are shown in \cref{fig:finite_relaxation_simulation}. We observe only a minor influence of $De$ on the pinch-off dynamics: stress relaxation further reduces the importance of elastic stress, which was already found to have a negligible influence on the process.

\subsection{2D Theory}
\subsubsection{Elastic stress}

We will now show why elasticity (in the dilute, Oldroyd-B limit) has little effect on bubble pinch-off, in contrast to the pinch-off of viscoelastic drops.
We note that the bubble neck is very slender close to pinch-off, which justifies a slender analysis of the problem.
This is in line with the quasi-two dimensional approach for bubble break-up in Newtonian fluids \cite{Gordillo.etal2005,Eggers.etal2007}.
Accordingly, we consider the collapse of cylindrical cavity for which we can analytically compute the elastic stress.
In what follows we will thus employ a two-dimensional approach, where any flow in the axial direction is neglected. The validity of this limitation will be tested, and confirmed, by comparing the 2D-model to the 3D numerical simulations that do take the axial flow into account. A further indication that axial flows are comparatively weak is obtained from \cref{app:flow} showing the motion of a small bubble that serves as a tracer particle.

We adopt a Lagrangian approach using the initial condition as a reference configuration, with radial coordinate \(R\).
The initial cavity radius corresponds to \(R=h_0\).
During the collapse, material points move to a current radial coordinate \(r\), according to a mapping \(r(R,t)\).
The position of the cavity radius is given by \(r=h(t)\).
Assuming a cylindrical symmetry, the mapping is dictated by mass conservation and takes the following form \cite{Green.Zerna1992,Carroll1987},
\begin{align}\label{eq:mapping}
  r^2(R,t) = R^2 + h^2(t) - h_0^2.
\end{align}
We now make use of the fact that without stress relaxation (\(\lambda=\infty\)), the conformation tensor \(\mathbf A\) follows the kinematics of the Finger tensor \cite{Snoeijer.etal2020}.
Hence, the conformation tensor becomes

\begin{align}
  A_{rr} = \left( \frac{\partial r
  }{\partial R}\right)^2=  \frac{R^2}{r^2}, \qquad A_{\theta\theta} = \frac{r^2}{R^2}.
\end{align}
In the expression for $A_{rr}$ we made use of volume conservation, which in this 2D model takes the form $\frac{\partial r}{\partial R} \frac{r}{R}=1$. 
Let us note that a correction on this expression for volume conservation, and thus also for $A_{rr}$, would appear in the presence of axial stretch. Making use of the mapping \eqref{eq:mapping}, the stress tensor can be written as
\begin{align}
  \frac{\sigma_{rr}}{G} = A_{rr} -1 = \frac{h_0^2 - h^2}{r^2}, \qquad \frac{\sigma_{\theta\theta}}{G} = A_{\theta\theta} -1 = \frac{h^2 - h_0^2}{r^2 + h_0^2 - h^2}.
  \label{eq:stress_theory}
\end{align}
These equations provide explicit expressions for the spatio-temporal evolution of the stress, for a given dynamics \(h(t)\).

\begin{figure}[tpb]
  \centering
  \includegraphics[width=17.2cm]{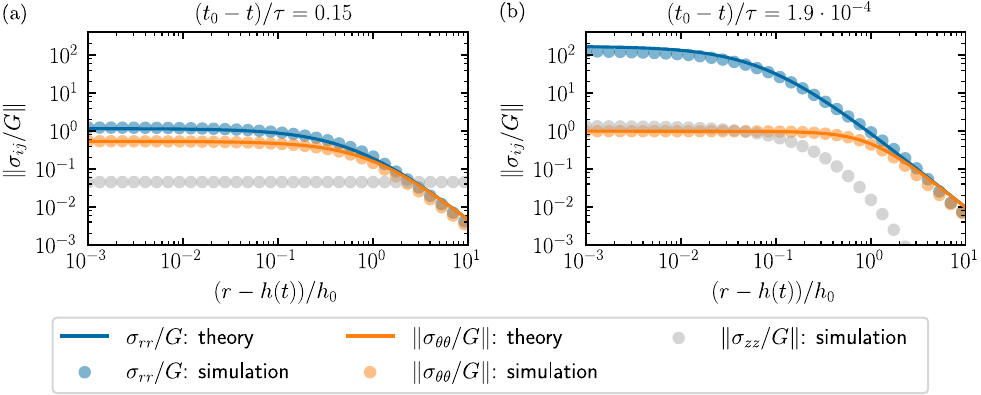}
  \caption{The fluid stresses for \(Ec = 0.1\) at different times plotted against the radial distance from the neck $r - h(t)$. (a) At early times the radial $\sigma_{rr}$ is of the same order as the azimuthal stress $\sigma_{\theta\theta}$, while the axial stress $\sigma_{zz}$ remains subdominant. (b) At later times close to breakup, the radial stress becomes much larger than the other two principal stress components.}
  \label{fig:stress_simulation}
\end{figure}

We now compare the stress predicted by the two-dimensional model to the results of our three-dimensional numerical simulations.
\cref{fig:stress_simulation} reports the elastic stress components in the fluid (\(Ec = 0.1\)) at \(t_0-t = 0.15\) and \(1.9\cdot 10^{-4}\), respectively, in panel (a) and (b).
Circles represent the numerical data, where the different colors correspond to $\sigma_{rr}$, $\sigma_{\theta\theta}$, and $\sigma_{zz}$. These numerical data can be used to test the 2D model, which neglected any axial flow. By consequence, the model only has predictions for $\sigma_{rr}$ and $\sigma_{\theta \theta}$, given by \cref{eq:stress_theory}. These are shown as the solid lines, and can be compared without any adjustable parameters to the 3D numerical results. It is clear that the model gives an excellent description of the radial and azimuthal stress, suggesting that the neglected axial flow is subdominant.
The results also show that close to \(t_0\), the stress in the radial direction dominates over the two other directions -- this observation formed the basis for the sketch proposed in \cref{fig:schematic_pinch-off}. 
It is of interest to evaluate the stress \cref{eq:stress_theory} inside the liquid at bubble edge (\(r=h\)), which close to pinch-off (\(h/h_0 \to 0\)) reduces to
\begin{align}
  \frac{\sigma_{rr}}{G} \simeq \left(\frac{h_0}{h}\right)^2, \qquad \frac{\sigma_{\theta\theta}}{G} \to -1.
\end{align} 
Hence, while the azimuthal stress remains finite, the radial stress tends to diverge (in line with \cref{fig:stress_simulation}b).

\subsubsection{Pinch-off}

The interesting situation is thus that the radial stress tends to diverge, but apparently this divergence does not inhibit the collapse of the neck.
To understand this, we first recall the
viscoelastic mechanics observed for droplet pinch-off. The momentum balance that dictates the dynamics of viscoelastic drops is in the axial direction, and involves the dominant component of stress $\sigma_{zz}$ (\cref{fig:schematic_pinch-off}). This axial stress scales as \(G(h_0/h)^4\) \cite{Clasen.etal2006}. 
Balancing this elastic stress with the capillary stress \(\gamma/h\), one obtains a typical neck size \(h \sim (Gh_0^4/\gamma)^{1/3}\), which indeed gives the scaling for the thread in \cref{fig:drop_bubble_contour_simulation} (bottom right panel).
The inclusion of stress relaxation (not present in our simulations) leads to a further exponential thinning set by the relaxation time.

Some important differences arise for bubble pinch-off. First, the dominant momentum balance is now in the radial direction, and involves $(\nabla \cdot \boldsymbol{\sigma})_r=\frac{1}{r}(\sigma_{rr}-\sigma_{\theta \theta})$. What we have learned from the model and simulations is that the dominant contribution comes from $\sigma_{rr} \sim G (h_0/h)^2$. While still singular as $h\to 0$, the divergence is significantly weaker than the elastic stress singularity observed during droplet pinch-off.
Second, the dynamics is inertial rather than capillary, so that the elastic stress must be compared to the inertial scale \(\rho \dot h^2\).
Ignoring the subtleties associated to logarithmic corrections, the inertial pinch-off follows \(h \sim (Bt)^{1/2}\), so that the inertial stress scales as \(\rho B^2/h^2\).
Hence, the peculiar situation arises that both elastic stress and inertial stress exhibit a divergence that scales as \(1/h^2\).
The fact that elasticity barely affects breakup suggests that elastic stress is typically much smaller than inertial stress, that is, \(G h_0^2\ll \rho B^2\).
This can indeed be understood from \cref{fig:concentration_simulation}a.
One observes that upon increasing \(G\) (via \(Ec\)), elasticity slows down the initial dynamics; in fact, for sufficiently large $G$ the capillary instability does not set in and break-up is fully inhibited.
Hence, for break-up to occur the capillary energy must be significantly larger than the elastic energy.
This capillary energy is converted into kinetic energy, which per unit axial length can be estimated as \(\sim \rho B^2\).
Since elastic energy was initially subdominant, it will remain subdominant during the entire pinch-off process.

\section{Conclusion}
\label{sec:Conclusion}
In this work, we have investigated the neck formation, thinning dynamics, and eventual break-up of the thread in the case of viscoelastic bubble pinch-off.
Experimentally, this was explored by changing the concentration, polymer length, and needle size.
Since the bubble and drop pinch-off processes exhibit qualitative similarities, we have compared the two phenomena in detail.
We have shown that there are some significant differences in thread formation and thinning dynamics.
The thread in bubble pinch-off is much thinner and lasts much shorter than the thread in drop pinch-off, and only forms at high concentrations when polymers start to overlap.
No threads were observed in the dilute limit.
In practice, the spatial resolution of our experiments is about 1 micron, and one might still wonder whether an even thinner thread might, in fact, have formed.
For this reason we performed numerical simulations and theoretical analysis using the Oldroyd-B model.
The theoretical results show that the stretching of the polymer is predominantly in the radial direction,  causing a much weaker divergence of the stress; this weaker singularity explains the absence of a thread in the dilute limit.

Previous work by Rajesh {\em et al.}~\cite{Rajesh.etal2022} already noticed that, in spite of qualitative similarities, viscoelastic air threads exhibit notable differences as compared to viscoelastic liquid threads.
In \cite{Rajesh.etal2022}, the thinning was fitted with an exponential law borrowed from drop pinch-off in the Oldroyd-B fluid.
As we have shown, such a law does not appear for bubble pinch-off in an Oldroyd-B fluid.
From an experimental point of view, some of our data might fit an exponential decay.
However, we do not find any evidence for such a universal thinning dynamics.
In particular, we find a dramatic effect of the needle size: thinner needles lead to significantly slower thinning dynamics.
A similar effect of drop size was reported for the break-up of extended capillary bridges \cite{Gaillard.etal2023}, but here the effect is much more pronounced, increasing the lifetime of a thread by orders of magnitude.
In addition, the nature of the instability and the potential formation of satellite bubbles is not universal and was found to depend on the size of the needle.

A central conclusion of our work is that viscoelastic bubble pinching is significantly different from the break-up of drops.
The capillary break-up of drops is routinely used as a rheological tool to extract the polymer relaxation time, for dilute suspensions.
We have demonstrated that for bubble pinch-off, the manifestations of viscoelasticity arise only beyond the dilute limit.
An important next step is to reproduce these manifestations using more advanced polymer models.
A potential outlook is that viscoelastic bubbles might offer a rheological probe that specifically targets properties at relatively high concentrations.

\section*{Acknowledgements}
This work was supported by an Industrial Partnership Program of the Netherlands Organization for Scientific Research (NWO), co-financed by Canon Production Printing Netherlands B.V., University of Twente, and Eindhoven University of Technology.
VS thanks Ayush Dixit for discussions and the initial version of the drop pinchoff code, and Saumili Jana for her work on AMR in \cite{sanjayElasticPinchOff}.
Furthermore, the simulation work was carried out on the national e-infrastructure of SURFsara, a subsidiary of SURF cooperation, the collaborative ICT organization for Dutch education and research. This simulation work was sponsored by NWO - Domain Science for the use of supercomputer facilities. The finite Deborah number cases have made use of the Hamilton HPC Service of Durham University.

\section*{Data Availability}
The experimental data that support the findings of this study are not publicly available due to the size of the dataset but are available from the corresponding author upon reasonable request. The Basilisk code used to perform the numerical simulations is freely available; see \cite{sanjayElasticPinchOff}.

\appendix
\section{Fluid flow around the bubble neck}
\label{app:flow}
\begin{figure}[htpb]
  \centering
  \includegraphics[width=0.8\textwidth]{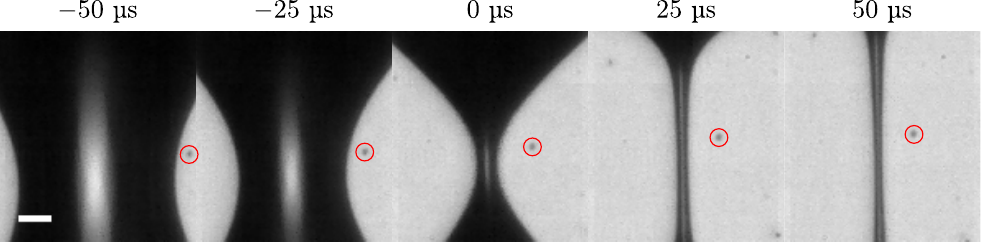}
  \caption{A series of snapshots of a micro bubble (highlighted by the red circles) close to the neck during the pinch-off at a concentration of \(1\, \textrm{wt.} \, \%\) PEO (\(M_W = 2.0 \times 10^6 \, \textrm{g/mol}\)) with \(d_{needle}=0.41\) mm. The scale bar is \(20\) \textmu m. (supplemental Movie S10 \cite{supplemental_material})}
  \label{fig:flow}
\end{figure}

It was not the intention of the experiments to measure the flow field around the bubble neck. However in some measurements we caught a micro bubble close to the neck during the pinch-off process. By tracking the position of the micro bubble, we can get an indication of the flow field around the neck, as shown in \cref{fig:flow}. This is not a replacement of a proper PIV measurement, but it does give some insight into the flow field, which is in line with our numerical simulations, the theory, and the hypothesis sketched in \cref{fig:schematic_pinch-off}: the flow is predominantly in the radial direction, and the axial flow is much smaller.

\section{Characterization of the polymer solutions}
Fluids exhibit several defining properties: viscosity, relaxation time, surface tension, and density.
To analyse the pinch-off process, we focus on viscosity and relaxation time, as they are most sensitive to polymer concentration and molecular length.
Although surface tension and density also vary with concentration, these changes are close to the experimental error of the measurement, so those are assumed to be constant.
Furthermore, since some polymer solutions are known to degrade over time, we repeated the fluid characterization on the first and last day of the experiments to check the degradation of the polymer solutions.

\subsection{Rheology}
\label{app:rheology}
\begin{figure}[htpb]
  \centering
  \includegraphics[width=\linewidth]{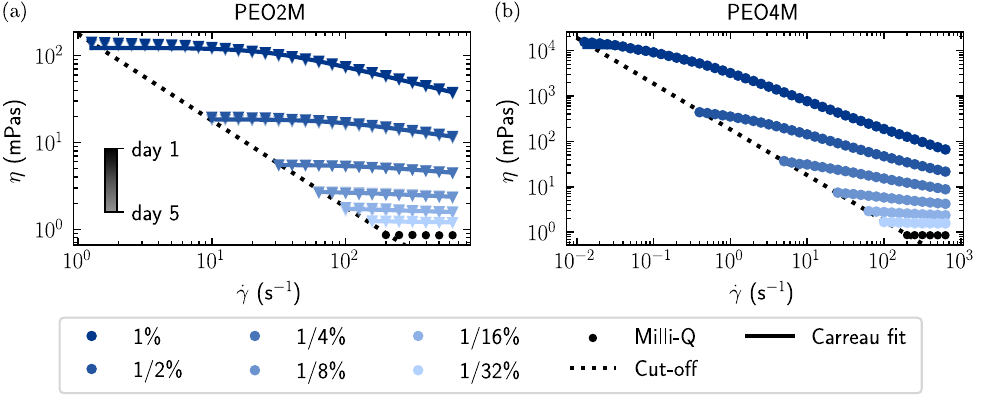}
  \caption{The viscosity of the polymer solutions as a function of the shear rate for different concentrations of PEO, \(M_W = 2.0 \times 10^6 \, \textrm{g/mol}\) (a) and \(4.0 \times 10^6 \, \textrm{g/mol}\) (b). The black dotted line indicates a measured torque value of \(5\cdot 10^{-3}\)~mN\(\cdot\)m, which is more than an order of magnitude above the rheometer's sensitivity due to the contribution of surface tension on the torque as is reported to be \(1\cdot 10^{-3}\)~mN\(\cdot\)m \cite{Johnston.Ewoldt2013}.}
  \label{fig:viscosity}
\end{figure}

The viscosity of the polymer solution is measured using a rheometer (Anton Paar, MCR 502) with a cone-plate geometry (CP50-1).
The viscosity is measured at room temperature as a function of the shear rate, which is increased from \(0.01 - 1000\, \textrm{s}^{-1}\) (see \cref{fig:viscosity}).
The black dotted line in \cref{fig:viscosity} indicates a measured torque value of \(5\cdot 10^{-3}\)~mN\(\cdot\)m. This value is more than an order of magnitude above the rheometer's sensitivity, which is around \(1\cdot 10^{-5}\)~mN\(\cdot\)m. However, since we are dealing with an aqueous fluid, it is not the torque sensitivity that limits our measurements, but rather the surface tension of the fluids that contributes to the torque on the rheometer. This surface tension torque is reported to be up to \(1\cdot 10^{-3}\)~mN\(\cdot\)m for water \cite{Johnston.Ewoldt2013}, which is below our cut-off value.
It has to be noted that the polymer solutions are shear thinning, and so viscosity is a function of the shear rate.
The viscosity is compared with the Carreau model, which estimates the polymer solutions' \(\mu_0\) and \(\mu_\infty\).
However, the results for \(\mu_\infty\) are not very reliable due to the extrapolation of the Carreau model.
Furthermore, the measurements are performed on the first and last day of the experiments, days 1 and 5, respectively.
As seen from \cref{fig:viscosity}, the viscosity of the polymer solutions is not significantly different between day 1 and day 5, so we can assume that the polymer solutions are not degraded during the experiments.

\subsection{Relaxation time}
\label{app:relaxation_time}
To determine the effective relaxation time (\(\lambda_d\)) of polymer solutions, we measured the neck width over time during pinch-off and fitted the results using the Oldroyd-B model:
\begin{align}
  \frac{h}{h_0} = \left(\frac{h_0 \mu}{\gamma}\right)^{1/3} e^{-\frac{t}{3\lambda_d}}.
\end{align}
Several methods can generate pinch-off, including drop pinch-off and CaBER.
We selected the drop pinch-off experiment due to its simplicity, as the neck forms naturally from a falling droplet under the influence of gravity.
Furthermore, it is the reversed counterpart of the bubble pinch-off experiment, which allows for a direct comparison.
A drawback of the drop pinch-off method is that we have to fit our measurements to a model to determine the relaxation time.
We have chosen the Oldroyd-B model since it is the simplest model that captures the exponential thinning of the thread, and is widely used in the literature to determine the relaxation time from drop pinch-off experiments.
However, it is important to note that the Oldroyd-B model is a simplified model that does not capture all the complexities of real polymer solutions, such as shear thinning and finite extensibility.
Therefore, the relaxation times determined from the drop pinch-off experiments should be considered as effective relaxation times that capture the dominant viscoelastic behavior of the polymer solutions in the context of our experiments.
Furthermore, our polymer solutions are created with a high molecular weight PEO, which is known to have a broad molecular weight distribution, resulting in a spectrum of relaxation times rather than a single one.

\begin{figure}[tpb]
  \centering
  \includegraphics[width=0.9\linewidth]{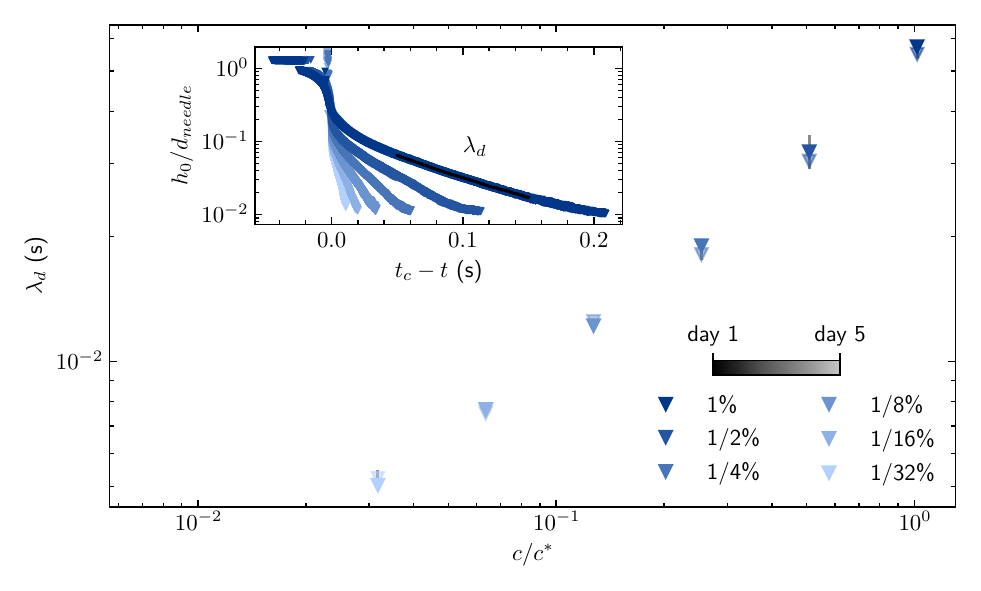}
  \caption{The relaxation time for PEO2M as a function of the concentration normalized by the overlap concentration.
  The opacity of the data points gives the day after preparation.
In the inset are the results of the drop experiments, where the width of the neck is given as function of time.
The relaxation time (\(\lambda_d\)) is determined by the slope of the straight line.}
  \label{fig:relaxation_timePEO2M}
\end{figure}

\begin{figure}[tpb]
  \centering
  \includegraphics[width=0.9\linewidth]{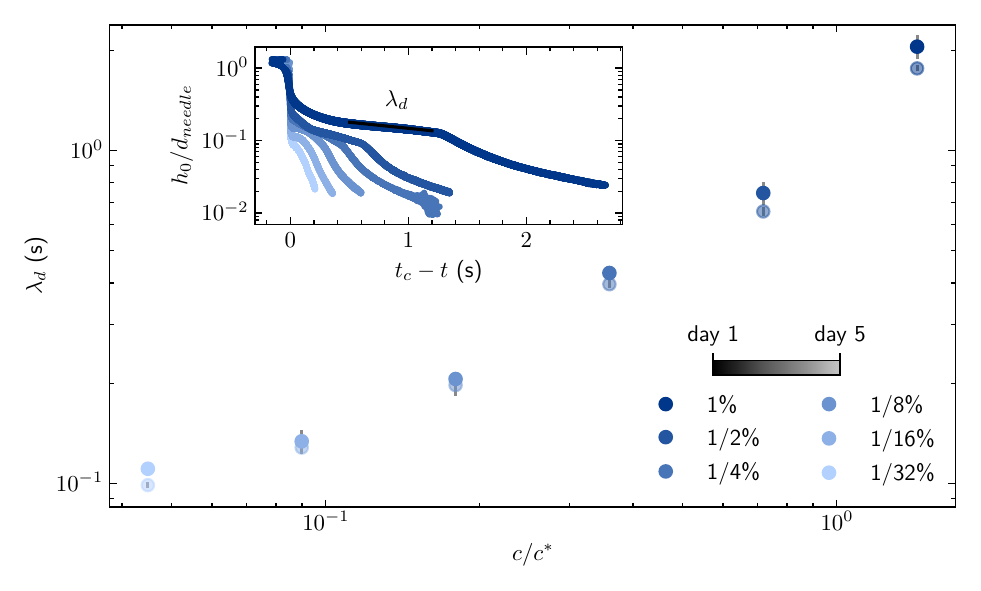}
\caption{The relaxation time for PEO4M as a function of the concentration normalized by the overlap concentration.
The opacity of the data points gives the day after preparation.
In the inset are the results of the drop experiments, where the width of the neck is given as function of time.
The relaxation time (\(\lambda_d\)) is determined by the slope of the straight line.}
  \label{fig:relaxation_timePEO4M}
\end{figure}

The inset of \cref{fig:relaxation_timePEO2M,fig:relaxation_timePEO4M} shows the neck width over time for different concentrations of PEO, \(M_W = 2.0 \times 10^6 \, \textrm{g/mol}\) and \(4.0 \times 10^6 \, \textrm{g/mol}\) respectively.
By fitting an exponential to the data's viscoelastic regime, we can determine the relaxation time given in the plots themselves.
We determined the effective relaxation time before and after the bubble pinch-off experiments to make sure that the effective relaxation time didn't change over the course of the experiments.
We see in \cref{fig:relaxation_timePEO2M,fig:relaxation_timePEO4M} that the relaxation time is not significantly different between days 1 and 5, so we can conclude that the polymer solutions are not degraded during the timespan of the experiments.

\clearpage

\bibliography{pinch_off}

@article{ingremeau2013stretching,
  author  = {Ingremeau, F. and Kellay, H.},
  title   = {Stretching Polymers in Droplet-Pinch-Off Experiments},
  journal = {Phys. Rev. X},
  volume  = {3},
  pages   = {041002},
  year    = {2013},
  doi     = {10.1103/PhysRevX.3.041002}
}

@article{dixit2025viscoelastic,
	title={Viscoelastic Worthington jets and droplets produced by bursting bubbles},
	author={Dixit, A. K. and Oratis, A. T. and Zinelis, K. and Lohse, D. and Sanjay, V.},
	journal={J. Fluid Mech.},
	volume={1010},
	pages={A2},
	doi = {10.1017/jfm.2025.237},
	year={2025}
}

@software{sanjayComphylabMultiRheoFlowV012025,
  title = {Comphy-Lab/{{MultiRheoFlow}}: V0.1},
  shorttitle = {Comphy-Lab/{{MultiRheoFlow}}},
  author = {Sanjay, V. and Collaborators},
  date = {2025-06-01},
  doi = {10.5281/ZENODO.15571585},
  organization = {Zenodo},
  version = {v0.1}
}

@software{sanjayComphylabElasticPinchOffInitial2025,
  title = {Comphy-Lab/{{ElasticPinchOff}}: Initial Release},
  shorttitle = {Comphy-Lab/{{ElasticPinchOff}}},
  author = {Sanjay, V.},
  date = {2025-06-01},
  doi = {10.5281/ZENODO.15571590},
  organization = {Zenodo},
  version = {v0.1}
}

@article{fattal2004constitutive,
	author = {Fattal, R. and Kupferman, R.},
	journal = {J. Non-Newtonian Fluid Mech.},
	number = {2-3},
	pages = {281--285},
	publisher = {Elsevier},
	title = {Constitutive laws for the matrix-logarithm of the conformation tensor},
	volume = {123},
	year = {2004}}

@article{lopez2019adaptive,
	author = {L{\'o}pez-Herrera, J.-M. and Popinet, S. and Castrej{\'o}n-Pita, A.-A.},
	journal = {J. Non-Newton. Fluid Mech.},
	pages = {144--158},
	publisher = {Elsevier},
	title = {An adaptive solver for viscoelastic incompressible two-phase problems applied to the study of the splashing of weakly viscoelastic droplets},
	volume = {264},
	year = {2019}}

@misc{basilliskpopinet,
	author = {Popinet, S. and {collaborators}},
	howpublished = {\url{http://basilisk.fr} (Last accessed: April, 2025)},
	title = {Basilisk {C}},
	year = {2013--2025}}

@misc{sanjayElasticPinchOff,
	author = {Sanjay, V. and {collaborators}},
	howpublished = {\url{https://github.com/comphy-lab/ElasticPinchOff} (Last accessed: April, 2026)},
	title = {Elastic{P}inch{O}ff},
	year = {2025--2026}}

@article{Amarouchene.etal2001,
  title = {Inhibition of the {{Finite-Time Singularity}} during {{Droplet Fission}} of a {{Polymeric Fluid}}},
  author = {Amarouchene, Y. and Bonn, D. and Meunier, J. and Kellay, H.},
  year = 2001,
  month = apr,
  journal = {Physical Review Letters},
  volume = {86},
  number = {16},
  pages = {3558--3561},
  issn = {0031-9007, 1079-7114},
  doi = {10.1103/PhysRevLett.86.3558},
  urldate = {2023-10-22}
}

@article{Anna.McKinley2001,
  title = {Elasto-{{Capillary Thinning}} and {{Breakup}} of {{Model Elastic Liquids}}},
  author = {Anna, Shelley L. and McKinley, Gareth H.},
  year = 2001,
  month = jan,
  journal = {Journal of Rheology},
  volume = {45},
  number = {1},
  pages = {115--138},
  issn = {0148-6055, 1520-8516},
  doi = {10.1122/1.1332389},
  urldate = {2023-01-09}
}

@article{Bazilevskii.etal1997,
  title = {Failure of {{Polymer Solution Filaments}}},
  author = {Bazilevskii, A V and Entov, V M and Lerner, M M and Rozhkov, A N},
  year = 1997,
  journal = {Polymer Science},
  volume = {39},
  number = {3},
  langid = {english}
}

@book{Bazilevsky.etal1990,
  title = {Liquid {{Filament Microrheometer}} and {{Some}} of {{Its Applications}}. {{In}}: {{Oliver}}, {{D}}.{{R}}. ({{Eds}}) {{Third European Rheology Conference}} and {{Golden Jubilee Meeting}} of the {{British Society}} of {{Rheology}}},
  author = {Bazilevsky, A.V. and Entov, V.M. and Rozhkov, A.N.},
  year = 1990,
  publisher = {Springer, Dordrecht},
  isbn = {978-94-010-6838-3}
}

@article{Bergmann.etal2006,
  title = {Giant {{Bubble Pinch-Off}}},
  author = {Bergmann, Raymond and Van Der Meer, Devaraj and Stijnman, Mark and Sandtke, Marijn and Prosperetti, Andrea and Lohse, Detlef},
  year = 2006,
  month = apr,
  journal = {Physical Review Letters},
  volume = {96},
  number = {15},
  pages = {154505},
  issn = {0031-9007, 1079-7114},
  doi = {10.1103/PhysRevLett.96.154505},
  urldate = {2025-09-22},
  copyright = {http://link.aps.org/licenses/aps-default-license},
  langid = {english}
}

@article{Burton.etal2005,
  title = {Scaling and {{Instabilities}} in {{Bubble Pinch-Off}}},
  author = {Burton, J. C. and Waldrep, R. and Taborek, P.},
  year = 2005,
  month = may,
  journal = {Physical Review Letters},
  volume = {94},
  number = {18},
  pages = {184502},
  issn = {0031-9007, 1079-7114},
  doi = {10.1103/PhysRevLett.94.184502},
  urldate = {2023-01-08}
}

@article{Carroll1987,
  title = {Pressure Maximum Behavior in Inflation of Incompressible Elastic Hollow Spheres and Cylinders},
  author = {Carroll, M. M.},
  year = 1987,
  month = apr,
  journal = {Quarterly of Applied Mathematics},
  volume = {45},
  number = {1},
  pages = {141--154},
  issn = {0033-569X, 1552-4485},
  doi = {10.1090/qam/885176},
  urldate = {2025-10-16},
  langid = {english}
}

@article{Clasen.etal2006,
  title = {The {{Beads-on-String Structure}} of {{Viscoelastic Threads}}},
  author = {Clasen, Christian and Eggers, Jens and Fontelos, Marco A. and Li, Jie and McKINLEY, Gareth H.},
  year = 2006,
  month = jun,
  journal = {Journal of Fluid Mechanics},
  volume = {556},
  pages = {283},
  issn = {0022-1120, 1469-7645},
  doi = {10.1017/S0022112006009633},
  urldate = {2023-01-09}
}

@article{Day.etal1998,
  title = {Self-{{Similar Capillary Pinchoff}} of an {{Inviscid Fluid}}},
  author = {Day, Richard F and Hinch, E John and Lister, John R},
  year = 1998,
  journal = {P H Y S I C A L R E V I E W L E T T E R S},
  volume = {80},
  number = {4},
  langid = {english}
}

@article{Deblais.etal2020,
  title = {Self-{{Similarity}} in the {{Breakup}} of {{Very Dilute Viscoelastic Solutions}}},
  author = {Deblais, A. and Herrada, M. A. and Eggers, J. and Bonn, D.},
  year = 2020,
  month = dec,
  journal = {Journal of Fluid Mechanics},
  volume = {904},
  pages = {R2},
  issn = {0022-1120, 1469-7645},
  doi = {10.1017/jfm.2020.765},
  urldate = {2023-10-22}
}

@book{deGennes1979,
  title = {Scaling {{Concepts}} in {{Polymer Physics}}},
  author = {{de Gennes}, P.G.},
  year = 1979,
  publisher = {Cornell University Press},
  isbn = {978-0-8014-1203-5},
  lccn = {lc78021314}
}

@article{Dekker.etal2022,
  title = {When {{Elasticity Affects Drop Coalescence}}},
  author = {Dekker, Pim J. and Hack, Michiel A. and Tewes, Walter and Datt, Charu and Bouillant, Ambre and Snoeijer, Jacco H.},
  year = 2022,
  month = jan,
  journal = {Physical Review Letters},
  volume = {128},
  number = {2},
  pages = {028004},
  issn = {0031-9007, 1079-7114},
  doi = {10.1103/PhysRevLett.128.028004},
  urldate = {2025-04-15},
  langid = {english}
}

@article{Eggers.etal2007,
  title = {Theory of the {{Collapsing Axisymmetric Cavity}}},
  author = {Eggers, J. and Fontelos, M. A. and Leppinen, D. and Snoeijer, J. H.},
  year = 2007,
  month = mar,
  journal = {Physical Review Letters},
  volume = {98},
  number = {9},
  pages = {094502},
  issn = {0031-9007, 1079-7114},
  doi = {10.1103/PhysRevLett.98.094502},
  urldate = {2023-01-11}
}

@article{Eggers.etal2020,
  title = {Self-{{Similar Breakup}} of {{Polymeric Threads}} as {{Described}} by the {{Oldroyd-B Model}}},
  author = {Eggers, J. and Herrada, M. A. and Snoeijer, J. H.},
  year = 2020,
  month = mar,
  journal = {Journal of Fluid Mechanics},
  volume = {887},
  eprint = {1905.12343},
  pages = {A19},
  issn = {0022-1120, 1469-7645},
  doi = {10.1017/jfm.2020.18},
  urldate = {2023-10-22},
  archiveprefix = {arXiv}
}

@book{Eggers.Fontelos2015,
  title = {Singularities: {{Formation}}, {{Structure}}, and {{Propagation}}},
  author = {Eggers, J. and Fontelos, M.A.},
  year = 2015,
  series = {Cambridge {{Texts}} in {{Applied Mathematics}}},
  publisher = {Cambridge University Press},
  isbn = {978-1-316-35239-7}
}

@article{Eggers1997,
  title = {Nonlinear Dynamics and Breakup of Free-Surface Flows},
  author = {Eggers, Jens},
  year = 1997,
  month = jul,
  journal = {Reviews of Modern Physics},
  volume = {69},
  number = {3},
  pages = {865--930},
  issn = {0034-6861, 1539-0756},
  doi = {10.1103/RevModPhys.69.865},
  urldate = {2025-10-21},
  copyright = {http://link.aps.org/licenses/aps-default-license},
  langid = {english}
}

@article{Entov.Hinch1997,
  title = {Effect of a {{Spectrum}} of {{Relaxation Times}} on the {{Capillary Thinning}} of a {{Filament}} of {{Elastic Liquid}}},
  author = {Entov, V.M. and Hinch, E.J.},
  year = 1997,
  month = sep,
  journal = {Journal of Non-Newtonian Fluid Mechanics},
  volume = {72},
  number = {1},
  pages = {31--53},
  issn = {03770257},
  doi = {10.1016/S0377-0257(97)00022-0},
  urldate = {2023-10-22}
}

@article{Gaillard.etal2022a,
  title = {What Determines the Drop Size in Sprays of Polymer Solutions?},
  author = {Gaillard, Antoine and Sijs, Rick and Bonn, Daniel},
  year = 2022,
  month = jul,
  journal = {Journal of Non-Newtonian Fluid Mechanics},
  volume = {305},
  pages = {104813},
  issn = {03770257},
  doi = {10.1016/j.jnnfm.2022.104813},
  urldate = {2025-11-05},
  langid = {english}
}

@misc{Gaillard.etal2023,
  title = {Beware of {{CaBER}}: {{Filament Thinning Rheometry Doesn}}'t {{Give}} `the' {{Relaxation Time}} of {{Polymer Solutions}}},
  shorttitle = {Beware of {{CaBER}}},
  author = {Gaillard, Antoine and Gutierrez, Miguel Angel Herrada and Deblais, Antoine and Eggers, Jens and Bonn, Daniel},
  year = 2023,
  month = sep,
  eprint = {2309.08440},
  urldate = {2023-10-18},
  archiveprefix = {arXiv},
  howpublished = {http://arxiv.org/abs/2309.08440}
}

@article{Gekle.etal2009,
  title = {Approach to Universality in Axisymmetric Bubble Pinch-Off},
  author = {Gekle, Stephan and Snoeijer, Jacco H. and Lohse, Detlef and Van Der Meer, Devaraj},
  year = 2009,
  month = sep,
  journal = {Physical Review E},
  volume = {80},
  number = {3},
  pages = {036305},
  issn = {1539-3755, 1550-2376},
  doi = {10.1103/PhysRevE.80.036305},
  urldate = {2024-10-04},
  copyright = {http://link.aps.org/licenses/aps-default-license},
  langid = {english}
}

@article{Gordillo.etal2005,
  title = {Axisymmetric {{Bubble Pinch-Off}} at {{High Reynolds Numbers}}},
  author = {Gordillo, J. M. and Sevilla, A. and {Rodr{\'i}guez-Rodr{\'i}guez}, J. and {Mart{\'i}nez-Baz{\'a}n}, C.},
  year = 2005,
  month = nov,
  journal = {Physical Review Letters},
  volume = {95},
  number = {19},
  pages = {194501},
  issn = {0031-9007, 1079-7114},
  doi = {10.1103/PhysRevLett.95.194501},
  urldate = {2023-10-18}
}

@book{Green.Zerna1992,
  title = {Theoretical {{Elasticity}}},
  author = {Green, A. E. and Zerna, W.},
  year = 1992,
  edition = {Second},
  publisher = {Dover Publications},
  isbn = {978-0-486-67076-8},
  langid = {english}
}

@article{Jiang.etal2017,
  title = {Bubble {{Pinch-off}} in {{Newtonian}} and {{Non-Newtonian Fluids}}},
  author = {Jiang, Xiao F. and Zhu, Chunying and Li, Huai Z.},
  year = 2017,
  month = oct,
  journal = {Chemical Engineering Science},
  volume = {170},
  pages = {98--104},
  issn = {00092509},
  doi = {10.1016/j.ces.2016.12.057},
  urldate = {2023-10-18}
}

@article{Johnston.Ewoldt2013,
  title = {Precision Rheometry: {{Surface}} Tension Effects on Low-Torque Measurements in Rotational Rheometers},
  shorttitle = {Precision Rheometry},
  author = {Johnston, Michael T. and Ewoldt, Randy H.},
  year = 2013,
  month = nov,
  journal = {Journal of Rheology},
  volume = {57},
  number = {6},
  pages = {1515--1532},
  issn = {0148-6055, 1520-8516},
  doi = {10.1122/1.4819914},
  urldate = {2026-05-21},
  langid = {english}
}

@article{Kawaguchi.etal1997,
  title = {Aqueous Solution Properties of Oligo- and Poly(Ethylene Oxide) by Static Light Scattering and Intrinsic Viscosity},
  author = {Kawaguchi, Seigou and Imai, Genji and Suzuki, Junto and Miyahara, Akira and Kitano, Toshiaki and Ito, Koichi},
  year = 1997,
  month = jun,
  journal = {Polymer},
  volume = {38},
  number = {12},
  pages = {2885--2891},
  issn = {00323861},
  doi = {10.1016/S0032-3861(96)00859-2},
  urldate = {2025-04-15},
  copyright = {https://www.elsevier.com/tdm/userlicense/1.0/},
  langid = {english}
}

@article{Keim.etal2006,
  title = {Breakup of {{Air Bubbles}} in {{Water}}: {{Memory}} and {{Breakdown}} of {{Cylindrical Symmetry}}},
  shorttitle = {Breakup of {{Air Bubbles}} in {{Water}}},
  author = {Keim, Nathan C. and M{\o}ller, Peder and Zhang, Wendy W. and Nagel, Sidney R.},
  year = 2006,
  month = oct,
  journal = {Physical Review Letters},
  volume = {97},
  number = {14},
  pages = {144503},
  issn = {0031-9007, 1079-7114},
  doi = {10.1103/PhysRevLett.97.144503},
  urldate = {2023-10-18}
}

@book{Larson1999,
  title = {The {{Structure}} and {{Rheology}} of {{Complex Fluids}}},
  author = {Larson, R.G.},
  year = 1999,
  series = {{{EngineeringPro Collection}}},
  publisher = {OUP USA},
  isbn = {978-0-19-512197-1},
  lccn = {98019940}
}

@article{Makhnenko.etal2021,
  title = {A Review of Liquid Sheet Breakup: {{Perspectives}} from Agricultural Sprays},
  shorttitle = {A Review of Liquid Sheet Breakup},
  author = {Makhnenko, Iaroslav and Alonzi, Elizabeth R. and Fredericks, Steven A. and Colby, Christine M. and Dutcher, Cari S.},
  year = 2021,
  month = sep,
  journal = {Journal of Aerosol Science},
  volume = {157},
  pages = {105805},
  issn = {00218502},
  doi = {10.1016/j.jaerosci.2021.105805},
  urldate = {2025-11-05},
  langid = {english}
}

@book{Morozov.Spagnolie2015,
  title = {Complex Fluids in Biological Systems: Experiment, Theory, and Computation},
  shorttitle = {Complex Fluids in Biological Systems},
  editor = {Morozov, Alexander and Spagnolie, Saverio E.},
  year = 2015,
  series = {Biological and Medical Physics, Biomedical Engineering},
  publisher = {Springer},
  address = {New York},
  isbn = {978-1-4939-2064-8 978-1-4939-2065-5},
  langid = {english}
}

@article{Morrison.Harlen2010,
  title = {Viscoelasticity in {{Inkjet Printing}}},
  author = {Morrison, Neil F. and Harlen, Oliver G.},
  year = 2010,
  month = jun,
  journal = {Rheologica Acta},
  volume = {49},
  number = {6},
  pages = {619--632},
  issn = {0035-4511, 1435-1528},
  doi = {10.1007/s00397-009-0419-z},
  urldate = {2023-10-17}
}

@article{Oratis.etal2023,
  title = {Coalescence of Bubbles in a Viscoelastic Liquid},
  author = {Oratis, Alexandros T. and Bertin, Vincent and Snoeijer, Jacco H.},
  year = 2023,
  month = aug,
  journal = {Physical Review Fluids},
  volume = {8},
  number = {8},
  pages = {083603},
  issn = {2469-990X},
  doi = {10.1103/PhysRevFluids.8.083603},
  urldate = {2025-04-09},
  langid = {english}
}

@article{Oratis.etal2024,
  title = {A Unifying {{Rayleigh-Plesset-type}} Equation for Bubbles in Viscoelastic Media},
  author = {Oratis, Alexandros T. and Dijs, Kay and Lajoinie, Guillaume and Versluis, Michel and Snoeijer, Jacco H.},
  year = 2024,
  month = feb,
  journal = {The Journal of the Acoustical Society of America},
  volume = {155},
  number = {2},
  pages = {1593--1605},
  issn = {0001-4966},
  doi = {10.1121/10.0024984},
  urldate = {2025-04-09},
  langid = {english}
}

@article{Papageorgiou1995,
  title = {On the Breakup of Viscous Liquid Threads},
  author = {Papageorgiou, Demetrios T.},
  year = 1995,
  month = jul,
  journal = {Physics of Fluids},
  volume = {7},
  number = {7},
  pages = {1529--1544},
  issn = {1070-6631, 1089-7666},
  doi = {10.1063/1.868540},
  urldate = {2025-10-21},
  langid = {english}
}

@article{Rajesh.etal2022,
  title = {Pinch-off of {{Bubbles}} in a {{Polymer Solution}}},
  author = {Rajesh, Sreeram and Peddada, Sumukh S. and Thi{\'e}venaz, Virgile and Sauret, Alban},
  year = 2022,
  month = dec,
  journal = {Journal of Non-Newtonian Fluid Mechanics},
  volume = {310},
  pages = {104921},
  issn = {03770257},
  doi = {10.1016/j.jnnfm.2022.104921},
  urldate = {2023-10-19}
}

@article{Sen.etal2021,
  title = {The {{Retraction}} of {{Jetted Slender Viscoelastic Liquid Filaments}}},
  author = {Sen, Uddalok and Datt, Charu and Segers, Tim and Wijshoff, Herman and Snoeijer, Jacco H. and Versluis, Michel and Lohse, Detlef},
  year = 2021,
  month = dec,
  journal = {Journal of Fluid Mechanics},
  volume = {929},
  pages = {A25},
  issn = {0022-1120, 1469-7645},
  doi = {10.1017/jfm.2021.855},
  urldate = {2023-10-17}
}

@article{Snoeijer.etal2020,
  title = {The {{Relationship}} between {{Viscoelasticity}} and {{Elasticity}}},
  author = {Snoeijer, J. H. and Pandey, A. and Herrada, M. A. and Eggers, J.},
  year = 2020,
  month = nov,
  journal = {Proceedings of the Royal Society A: Mathematical, Physical and Engineering Sciences},
  volume = {476},
  number = {2243},
  pages = {20200419},
  issn = {1364-5021, 1471-2946},
  doi = {10.1098/rspa.2020.0419},
  urldate = {2023-04-25}
}

@book{Tanner2000,
  title = {Engineering {{Rheology}}},
  author = {Tanner, R.I.},
  year = 2000,
  series = {Oxford {{Engineering Science Series}}},
  publisher = {OUP Oxford},
  isbn = {978-0-19-159016-0}
}

@article{Thoroddsen.etal2007,
  title = {Experiments on {{Bubble Pinch-Off}}},
  author = {Thoroddsen, S. T. and Etoh, T. G. and Takehara, K.},
  year = 2007,
  month = apr,
  journal = {Physics of Fluids},
  volume = {19},
  number = {4},
  pages = {042101},
  issn = {1070-6631, 1089-7666},
  doi = {10.1063/1.2710269},
  urldate = {2023-01-09}
}

@misc{supplemental_material,
  note = {See Supplemental Material at [URL will be inserted by publisher]
          for additional details and Movies S1-10}
}

@phdthesis{Stokes1998,
  author       = {J. R. Stokes},
  title        = {Swirling flow of viscoelastic fluids},
  school       = {The University of Melbourne},
  address      = {Melbourne, Australia},
  type         = {PhD thesis},
  note         = {Department of Chemical Engineering},
  year         = {1998},
}

@article{Brackbill.etal1992,
  author = {Brackbill, J. U. and Kothe, D. B. and Zemach, C.},
  title = {A continuum method for modeling surface tension},
  journal = {Journal of Computational Physics},
  volume = {100},
  number = {2},
  pages = {335--354},
  year = {1992},
  doi = {10.1016/0021-9991(92)90240-Y}
}

@article{Popinet2009,
  author = {Popinet, S.},
  title = {An accurate adaptive solver for surface-tension-driven interfacial flows},
  journal = {Journal of Computational Physics},
  volume = {228},
  number = {16},
  pages = {5838--5866},
  year = {2009},
  doi = {10.1016/j.jcp.2009.04.042}
}

\end{document}